\documentclass[pra,amsmath,amssymb,amsfonts,twocolumn,superscriptaddress]{revtex4}

\usepackage{graphicx}
\usepackage{dcolumn}
\usepackage{bm}
\usepackage{bbm}

\begin{document}

\preprint{APS/123-QED}

\title{Deterministic multi-mode photonic device for quantum information processing}

\author{Anne E. B. Nielsen}
\author{Klaus M{\o}lmer}
\affiliation{Lundbeck Foundation Theoretical Center for Quantum
System Research, Department of Physics and Astronomy,
Aarhus University, DK-8000 {\AA}rhus C, Denmark}

\date{\today}

\begin{abstract}
We propose the implementation of a light source, which can deterministically generate a rich variety of multi-mode quantum states. The desired states are encoded in the collective population of different ground hyperfine states of an atomic ensemble and converted to multi-mode photonic states by excitation to optically excited levels followed by cooperative spontaneous emission. Among our examples of applications, we demonstrate how two-photon entangled states can be prepared and implemented in a protocol for reference frame free quantum key distribution and how one-dimensional as well as higher-dimensional cluster states can be produced.
\end{abstract}

\pacs{42.50.Dv, 42.50.Ex, 42.50.Nn, 32.80.Ee}
\keywords{Suggested keywords}

\maketitle

\section{Introduction}

Quantum states of light are efficient carriers of quantum information with applications in quantum computing, quantum cryptography, and quantum networks. Their most serious drawback is the absence of suitable non-linear interactions which can produce and transform the desired states. Measurement back action presents one means to accommodate the required non-linearities and paves the way for universal quantum computing with the tools of linear optics, but certain resource states such as single-photon states \cite{knill} or superpositions of coherent states \cite{ralph} are required, while even more involved states, such as cluster states \cite{cluster}, are needed to perform successfully with high probability. An efficient method to prepare general quantum states of light deterministically would constitute a significant step forward for optical quantum information processing.

There are a number of different strategies to produce non-classical states of light. Squeezed states and multi-mode states with quantum correlated amplitudes can be produced directly by optical parametric oscillators and beam splitters, while these states can be further transformed into states with non-Gaussian amplitude distributions by conditional measurements \cite{wakui,neergaard,grangier}. The probabilistic nature of the schemes, however, limits the variety of states that can be produced by this method. Another successful approach uses laser excitation of an atomic ensemble with all atoms in the same single atom state and detection of a spontaneously emitted photon to conditionally prepare a superposition state with different atoms populating the final state of the spontaneous emission process. The atomic state is generated at a random instant and the ensemble is then ready to deterministically emit a single photon at the moment when the experimentalist couples the singly occupied state to an optically excited state \cite{DLCZ,expDLCZ,spsource}.

\begin{figure}
\includegraphics[width=\columnwidth]{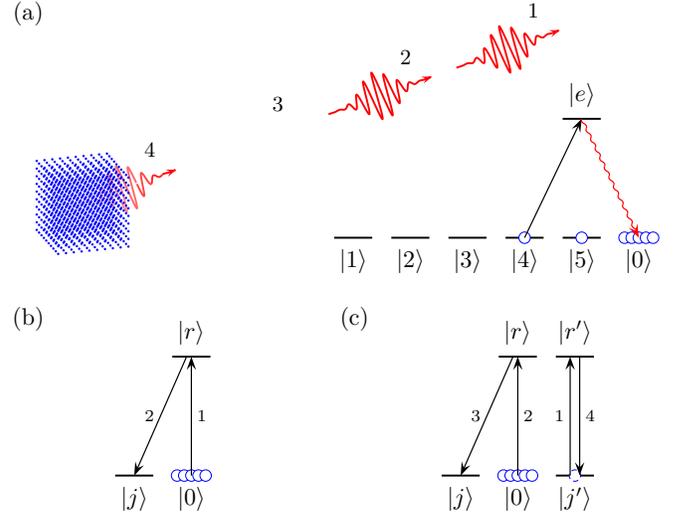}
\caption{(Color online) (a) Step-wise transfer of a quantum state encoded in a cold atomic ensemble to light. The initial state of the ensemble is an arbitrary superposition of states with zero or one atom in the single-atom levels $|1\rangle$ to $|5\rangle$, and during the conversion process, occupation by one atom translates into a single photon in a corresponding spatio-temporal light mode. For the case illustrated, the states $|1\rangle$, $|2\rangle$, $|4\rangle$, and $|5\rangle$ are occupied initially. (b) Preparation of the atomic state. The excitation from $|0\rangle$ to a Rydberg state $|r\rangle$ is off-resonant if more than a single atom is excited, and a collectively distributed single excitation can be subsequently transferred to a unit atomic population of the single-atom ground state level $|j\rangle$. (c) The feeding of state $|j\rangle$ by the attempted transfer $|0\rangle \rightarrow |r\rangle \rightarrow |j\rangle$ can also be made conditional on the population of another state $|j'\rangle$ if that state is excited to a Rydberg state $|r'\rangle$ interacting with $|r\rangle$.}\label{convert}
\end{figure}

In the present paper, we propose a source, which is capable of producing a large variety of quantum states of light. We focus on coding of quantum bits for quantum computing and communication, and we use multiple spatial and temporal modes to provide multi-bit registers with different entangled states. The basic idea is to prepare the equivalent state in an ensemble of cold atoms and, in a step-wise process, convert the atomic state to a quantum state of light, Fig.~\ref{convert}(a), as proposed for two-mode states in \cite{cirac}. To avoid the requirement of individual addressing of the atoms and to facilitate the efficient conversion to light, we encode the quantum state in the collective populations of different internal states of the atoms as proposed in \cite{collective}. The transfer of precisely one atom in the ensemble to an initially unoccupied internal state is carried out via intermediate excitation of a Rydberg excited state and makes use of the so-called Rydberg blockade mechanism \cite{blockade,lukin}. It is thus possible to initialize the ensemble with all atoms populating a definite internal "reservoir" state, denoted $|0\rangle$ in the following, and to subsequently encode, in principle, any quantum superposition states of an $N$-bit register based on $N$ different internal states, labeled $|1\rangle,|2\rangle,\ldots,|N\rangle$, with each qubit value set to zero if the state is unpopulated, and to unity, if the corresponding state is populated by precisely one atom. This is ensured by the universality of the collective quantum computing proposal with Rydberg blockade \cite{collective}. Here, however, we also have the freedom to process the state of the atomic ensemble between light emission events. We can thus reuse the atomic levels and generate entangled light states with more than $N$ qubits. Our current task with the Rydberg blocked ensemble is a more dedicated one than quantum computing, and rather than appeal to universality, we will describe explicit schemes which in few, efficient steps use the Rydberg interaction, the physical properties of the internal atomic level structure, and the collective emission properties of the ensemble to provide the optical states of interest.

\section{State preparation and emission}

The Rydberg blockade mechanism \cite{blockade} utilizes the large dipole-dipole interaction between atoms in certain Rydberg states that are not separated by more than about $10~\mu$m. Due to this interaction, states with two or more atoms in Rydberg excited states have energies that are significantly different from the sum of the single atom excitation energies, and therefore resonant excitation of a few $\mu\textrm{m}$ sized ensemble from the reservoir ground atomic state will coherently drive the system towards the state with a single coherently distributed Rydberg excitation in the ensemble. The Rabi frequency for this transition is collectively enhanced by the factor $\sqrt{K}$, where $K$ is the number of atoms. This enhancement and the blocking of further excitation has been seen for a pair of atoms $(K=2)$ \cite{gatePG,blockMS}, and blocking and enhanced coupling has also been observed in larger samples of atoms \cite{Pfau}. After a $\pi$ pulse on the coherent transition, there is a single ensemble Rydberg excitation and one can subsequently drive this collectively shared population into a low lying state, Fig.~\ref{convert}(b). To provide the coupling in a fully selective manner between the desired internal states, we suggest to use polarization selection rules and the optical frequency resolution of different Zeeman ground states of the atoms in a homogeneous magnetic field switched on during state selective processes.

In the following, the ket $|n_1n_2\ldots n_N\rangle$ denotes a state with $n_j\in\{0,1\}$ atoms in the single-atom state $|j\rangle$, $j=1,2,\ldots,N$, and the remaining atoms in the reservoir state $|0\rangle$. In writing this state, phase factors arising due to the different positions of the atoms within the laser traveling wave fields are taken into account. The state $|10\ldots0\rangle$, for instance, with one atom in $|1\rangle$ and zero atoms in $|2\rangle$ to $|N\rangle$ is
\begin{equation}\label{state}
|10\ldots0\rangle=\frac{1}{\sqrt{K}}\sum_{j=1}^Ke^{i\mathbf{k}_0\cdot\mathbf{r}_j}
|0\rangle_1|0\rangle_2\cdots|1\rangle_j\cdots|0\rangle_K,
\end{equation}
where $K\gg N$ is the number of atoms in the ensemble, $|j\rangle_k$ means that the $k$th atom is in the internal state $|j\rangle$, $\mathbf{r}_j$ is the position of the $j$th atom, and $\mathbf{k}_0$ is the resulting wave vector of the light pulses used to bring the ensemble to the above state, i.e., $\mathbf{k}_0=\sum_i \delta n_i\mathbf{k}_i$, where $\mathbf{k}_i$ is the wave vector of the $i$th laser pulse and $\delta n_i$ is $1$ if a photon is absorbed and $-1$ if a photon is emitted during the $i$th pulse.

If we apply a $\pi$-pulse with wave vector $\mathbf{k}_e$ to transfer the atom in the state $|1\rangle$ to the optically excited state $|e\rangle$, the amplitude for emission of a photon with wave vector $\mathbf{k}$ and polarization $\mathbf{e}_{\mathbf{k}}$ when the atom decays to the state $|0\rangle$ is proportional to $K^{-1/2}\sum_{j=1}^K
\langle0|\mathbf{e}_{\mathbf{k}}\cdot\mathbf{d}|e\rangle
e^{i(\mathbf{k}'_0-\mathbf{k})\cdot\mathbf{r}_j}$,
where $\mathbf{d}$ is the dipole operator and $\mathbf{k}'_0=\mathbf{k}_0+\mathbf{k}_e$. The transition probability $P(\mathbf{k})$ is thus proportional to
\begin{equation}
P(\mathbf{k})\propto\frac{1}{K}\left|\sum_{j=1}^K
e^{i(\mathbf{k}'_0-\mathbf{k})\cdot\mathbf{r}_j}\right|^2.
\end{equation}
The factor $1/K$ suppresses emission in all directions except $\mathbf{k}\approx\mathbf{k}'_0$, for which the sum is approximately $K$ and $P(\mathbf{k})\propto K$. $c|\mathbf{k}|$ must also equal the atomic excitation frequency. The factor $1/K$ is replaced by $1/K^2$ if the atom decays to a different hyperfine level, and such transitions are thus also suppressed. With an ensemble of about $1000$ atoms it is possible both to satisfy the requirement of all atoms being at most $10\textrm{ }\mu\textrm{m}$ apart, while the ensemble is still large enough to ensure directional emission of the photon \cite{mark,line}. A detailed analysis of light emission from various arrangements of atoms can be found in \cite{cirac,porras}. To map a general superposition of the states $|n_1n_2\ldots n_N\rangle$ to $N$ temporally distinct light pulses, we simply transfer one internal state population after the other to the optically excited state as illustrated schematically for $N=5$ in Fig.~\ref{convert}(a), and we can even renew the populations of the atomic states during the conversion process if desired.

\section{GHZ and Bell states}

We now turn to a description of explicit schemes to generate particular quantum states of light. We start with the $M$-mode GHZ state $\left(|+\rangle_1|+\rangle_2\ldots|+\rangle_M+ |-\rangle_1|-\rangle_2\ldots|-\rangle_M\right)/\sqrt{2}$, where $|+/-\rangle_m$ denotes a single/zero-photon state in the $m$th mode. To prepare this state, we first apply a $\pi/2$-pulse between $|0\rangle$ and a Rydberg level $|r'\rangle$ followed by a $\pi$-pulse between $|r'\rangle$ and the ground state level $|j'\rangle$, which leads to a superposition of having zero and one atom in $|j'\rangle$. A train of entangled pulses, all containing either zero or one photon, can then be emitted by conditional feeding of the state $|j\rangle$, as shown in Fig.~\ref{convert}(c). After $M-1$ emissions, we apply a $\pi$-pulse between $|j'\rangle$ and $|r'\rangle$, a $\pi$-pulse between $|r'\rangle$ and $|0\rangle$, and a $\pi$-pulse between $|j'\rangle$ and $|r'\rangle$, which coherently changes the population in $|j'\rangle$ between zero and one. The last mode is then released by transferring the population in $|j'\rangle$ to an excited state.

\begin{figure}
\includegraphics[width=\columnwidth]{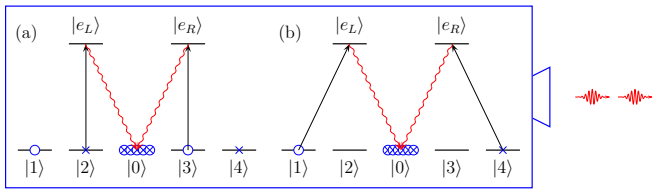}
\caption{(Color online) Generation of the photonic Bell state $(|R\rangle_1|L\rangle_2-|L\rangle_1|R\rangle_2)/\sqrt{2}$ from the collective atomic state $(|1010\rangle-|0101\rangle)/\sqrt{2}$ (the first term is indicated with circles, and the second with crosses). (a) A light pulse transfers the populations in $|2\rangle$ and $|3\rangle$ to the excited states $|e_L\rangle$ and $|e_R\rangle$, respectively, and the subsequent atomic decay to the state $|0\rangle$ leads to emission of the first photon, which is $L$-polarized if $|e_L\rangle$ is populated and $R$-polarized if $|e_R\rangle$ is populated. (b) The second photon is similarly obtained via spontaneous emission after transfer of the populations in the states $|1\rangle$ and $|4\rangle$ to the excited states.}\label{englert}
\end{figure}

The maximally entangled two-mode Bell state $|\Psi_-\rangle=(|R\rangle_1|L\rangle_2-|L\rangle_1|R\rangle_2)/\sqrt{2}$, where $R$ and $L$ denote right and left polarized photons, respectively, can be prepared, if we first prepare the atoms in the collective state $|\Psi_-^\textrm{at}\rangle=(|1010\rangle-|0101\rangle)/\sqrt{2}$, making use of the five hyperfine states in Fig.~\ref{englert}. This is done as follows: \textit{i}) Transfer one atom from $|0\rangle$ to $|2\rangle$ via a Rydberg level. \textit{ii}) Use a Raman transition between $|2\rangle$ and $|3\rangle$ to obtain the transformation $|0100\rangle\rightarrow(|0010\rangle-|0100\rangle)/\sqrt{2}$. \textit{iii}) Move an atom from $|0\rangle$ to $|4\rangle$ if and only if $|3\rangle$ is unoccupied, and an atom from $|0\rangle$ to $|1\rangle$ if and only if $|2\rangle$ is unoccupied using the conditional method in Fig.~\ref{convert}(c). The optical release of the state $|\Psi_-^\textrm{at}\rangle$ is then achieved as explained in Fig.~\ref{englert}. If the light pulses are released by spontaneous Raman transitions in the atoms, the temporal shape of the emitted photon pulse can be controlled by the coupling laser amplitude. Conversely, incident light pulses with a known temporal shape can be transferred via Raman processes to collective atomic population in different states. By subsequent application of Rydberg quantum gates, the state can be processed and analyzed, and the ensemble can thus, for example, be used to distinguish between all four polarization or photon number encoded Bell states.

As a specific application of the two-mode optical Bell state, we implement the double trine scheme for quantum key distribution, proposed recently by Tabia and Englert \cite{englert}. This scheme has the appealing features that the users of the communication channel, Alice and Bob, do not need to share a common reference frame and it is possible to extract $0.573$ key bits per trine state which is close the the theoretical maximum. The scheme applies a sequence of groups of three photons with orthogonal polarizations $R$ and $L$. The groups are chosen at random among three mixed quantum states $\rho_i$, $i=1,2,3$, each being a direct product of a completely mixed state $(|R\rangle\langle R|+|L\rangle\langle L|)/2$ of the $i$th photon and a pure Bell state $(|R\rangle|L\rangle-|L\rangle|R\rangle)/\sqrt{2}$ of the other two photons. To produce such a state, we prepare the atomic state $|\Psi_-^\textrm{at}\rangle$ and interweave its conversion to light with the emission of a mixed polarization state of a single photon. The mixed state $(|R\rangle\langle R|+|L\rangle\langle L|)/2$ can be generated by using a classical random decision to determine whether an atom is moved from $|0\rangle$ to $|e_L\rangle$ or from $|0\rangle$ to $|e_R\rangle$ via a Rydberg level. The subsequent decay back to $|0\rangle$ then produces the photon with the apparent random polarization. Note that the emission of the randomly polarized photon may occur independently of the state of the levels $|1\rangle$, $|2\rangle$, $|3\rangle$, and $|4\rangle$, and the three photons can thus be emitted in any order, which allows Alice to prepare any of the states $\rho_i$. The detection by Bob consists simply in measuring the polarization of the individual photons in any basis and does not require any further complicated processing of the states.

\begin{figure}
\includegraphics[width=\columnwidth]{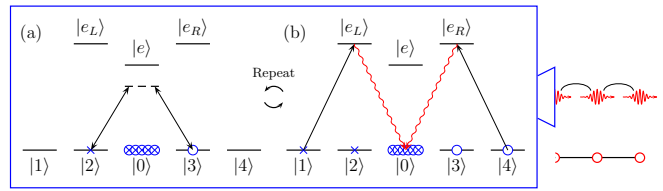}
\caption{(Color online) Generation of a one-dimensional cluster state. Initially, one atom is moved to $|3\rangle$. (a) In each iteration step, a $\pi/2$-pulse is applied to the (Raman) transition between $|2\rangle$ and $|3\rangle$. An atom is then moved from $|0\rangle$ to $|1\rangle$ if and only if $|3\rangle$ is unoccupied, and an atom is moved from $|0\rangle$ to $|4\rangle$ if and only if $|2\rangle$ is unoccupied. (Note that the sum of the populations in $|2\rangle$ and $|3\rangle$ is always one atom.) (b) The populations in $|1\rangle$ and $|4\rangle$ are transferred to $|e_L\rangle$ and $|e_R\rangle$, respectively, which leads to emission of a photon, and the process is repeated. The bonds between the emitted photons illustrate the cluster state entanglement, and the graph representing the state is shown below the photons.}\label{rudolph}
\end{figure}

\section{Cluster states}

Recently, Lindner and Rudolph proposed to use a quantum dot to prepare one-dimensional cluster states of photons \cite{rudolph,twod}. Following their approach, such cluster states can be produced as shown in Fig.~\ref{rudolph}. Starting with all atoms in $|0\rangle$, the initialization consists in moving one atom to the state $|3\rangle$ via a Rydberg level. We then use a Raman transition to apply a $\pi/2$ pulse between $|2\rangle$ and $|3\rangle$. Using the controlled feeding of atomic states shown in Fig.~\ref{convert}(c), we can apply the transformations $|0100\rangle\rightarrow|1100\rangle$ and $|0010\rangle\rightarrow|0011\rangle$ and subsequently transfer the populations in $|1\rangle$ and $|4\rangle$ to $|e_L\rangle$ and $|e_R\rangle$, respectively. After emission of a photon thus entangled with the atomic population remaining in states $|2\rangle$ and $|3\rangle$, one applies again a $\pi/2$-pulse between $|2\rangle$ and $|3\rangle$ followed by a new Rydberg controlled feeding of the populations in $|1\rangle$ and $|4\rangle$ and emission of the next photon. This procedure is iterated and one gradually builds up a one-dimensional cluster state of light pulses. After the desired number of iterations, the cluster is decoupled from the atomic state by measuring the state of the last emitted photon in the basis $\{|L\rangle,|R\rangle\}$.

\begin{figure}
\includegraphics[width=\columnwidth]{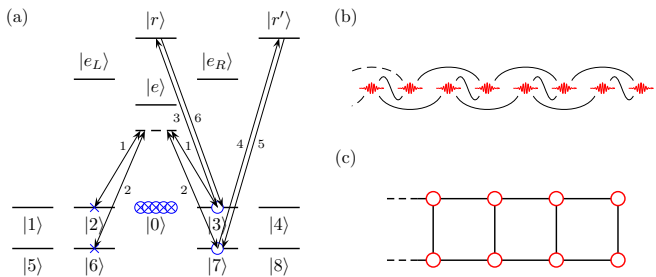}
\caption{(Color online) Generation of a two-dimensional cluster state. Initially, one atom is moved to $|3\rangle$, and one atom is moved to $|7\rangle$. (a) In each iteration step, a $\pi/2$-pulse is applied to the transition between $|2\rangle$ and $|3\rangle$ (pulse 1) and to the transition between $|6\rangle$ and $|7\rangle$ (pulse 2). The $\pi$-pulses 3-6 then change the phase of the state by $\pi$ unless $|3\rangle$ and $|7\rangle$ are both unoccupied. Atoms are moved from $|0\rangle$ to $|1\rangle$, $|4\rangle$, $|5\rangle$, and $|8\rangle$ conditioned on $|3\rangle$, $|2\rangle$, $|7\rangle$, and $|6\rangle$ being unoccupied, respectively, and photons are emitted by transferring first the populations in $|1\rangle$ and $|4\rangle$ and subsequently the populations in $|5\rangle$ and $|8\rangle$ to $|e_L\rangle$ and $|e_R\rangle$. (b) The first 8 emitted photons and (c) the corresponding graph.}\label{rudolph2}
\end{figure}

The behavior of one-dimensional cluster states can be efficiently simulated on a classical computer, and they are hence not useful resource states for cluster state quantum computing \cite{nielsen}. Our method can, however, be generalized to higher-dimensional cluster states by involving more internal levels of the atoms. To generate, for instance, the two-dimensional cluster state shown in Fig.~\ref{rudolph2}(c), we can use eight hyperfine states to prepare two one-dimensional cluster states in parallel and then add a controlled phase gate in each iteration step to obtain the bonds between them. The details of the processing of the state of the ensemble between each emission of photons are shown in Fig.~\ref{rudolph2}(a), and the emission follows the same procedure as for the one-dimensional cluster state. The cluster state is decoupled from the atomic ensemble by measuring the state of the last two emitted photons. Larger two-dimensional cluster states can be obtained by weaving more one-dimensional cluster states together, and one can also form tubes and three-dimensional structures. In reality, the finite number of internal states of the atoms restricts the possibilities, and to comply with this, we suggest to emit the light pulses from a one- or two-dimensional array of individually addressable ensembles with separations that are small enough to allow long distance Rydberg phase gates \cite{arrayMS} between nearest neighbor ensembles.

\section{Conclusion}

In conclusion, we have proposed a multi-purpose, multi-mode photonic source and constructed explicit schemes to prepare GHZ states, double trine states, and cluster states. We have only considered states with at most one photon per mode, but we anticipate that the same mechanism allows emission of more photons into the same mode \cite{cirac}. The collective encoding generalizes naturally to this case by allowing atomic state populations exceeding one. Experiments on Rydberg mediated one- and two-bit operations are still in their infancy, but according to theoretical estimates, their fidelities may be high enough that entangled photonic states with tens of modes may be prepared. A Rydberg state mediated quantum gate may typically be accomplished in about $10^{-6}$~s, and recent experimental results with pulsed laser fields even demonstrate Rabioscillations to Rydberg states taking place on a time scale of about $1$~ns \cite{pc}. The time required to emit a single photon can be as short as $10^{-7}$~s \cite{line}. These time scales are much faster than Rydberg state lifetimes on the order of milliseconds. The shape of the laser pulses used to excite the atoms leaves some freedom to tailor the shape and duration of the emitted pulses, and the emission direction is controllable and may vary from mode to mode. The number of photons that can be emitted by a single ensemble is limited in practice, because the interactions with the ensemble are not perfectly symmetric. The collective state of the ensemble may, however, be brought back to the symmetric state by optical pumping of all atoms into the reservoir state or by a transfer of the atomic internal state populations via the Rydberg gate mechanism to a new ensemble from which the photon emission can proceed.

\begin{acknowledgments}
This work was supported by ARO-IARPA.
\end{acknowledgments}

\end{document}